\documentclass[reprint,twocolumn,pre,showpacs,amsmath,amssymb,aps]{revtex4-1}
\usepackage{natbib}	
\usepackage{graphicx,color,rotating}
\usepackage[latin1]{inputenc}
\usepackage{textcomp}
\usepackage{dcolumn}% Align table columns on decimal point
\usepackage{bm}     % bold math
\usepackage{upgreek}
\usepackage{subdepth}  			% Added by JTK (aligns sub-indices)
\usepackage[latin1]{inputenc}

\usepackage{array}
\newcolumntype{L}[1]{>{\raggedright\let\newline\\\arraybackslash\hspace{0pt}}m{#1}}
\newcolumntype{C}[1]{>{\centering\let\newline\\\arraybackslash\hspace{0pt}}m{#1}}
\newcolumntype{R}[1]{>{\raggedleft\let\newline\\\arraybackslash\hspace{0pt}}m{#1}}
\renewcommand{\vec}[1]{\bm{#1}}
\newcommand{\ee}{\mathrm{e}}
\newcommand{\ii}{\mathrm{i}}

\newcommand{\pp}{\partial}

\newcommand{\nablabf}{\boldsymbol{\nabla}}

\newcommand{\divop}{\nablabf\cdot}

\newcommand{\fffac}{\vec{f}_\mathrm{ac}}
\newcommand{\fffacI}{\vec{f}_\mathrm{ac}^{(1)}}

\newcommand{\gvec}{\vec{g}}

\newcommand{\nnn}{\vec{n}}

\newcommand{\rrr}{\vec{r}}

\newcommand{\vvv}{\vec{v}}

\newcommand{\cO}{c_0}

\newcommand{\EacO}{E_\mathrm{ac}^{(0)}}

\newcommand{\etaO}{\eta_0}
\newcommand{\etaOO}{\eta_0^{(0)}}

\newcommand{\rhat}{{\hat{r}{}}}

\newcommand{\yhat}{{\hat{y}{}}}
\newcommand{\zhat}{{\hat{z}{}}}

\newcommand{\kOO}{k_0^{(0)}}

\newcommand{\kapO}{\kappa_0}
\newcommand{\kapOO}{\kappa_0^{(0)}}

\newcommand{\pa}{p_\mathrm{a}}

\newcommand{\pI}{p_1}
\newcommand{\pIO}{p_1^{(0)}}
\newcommand{\pIOh}{\hat{p}_1^{(0)}}

\newcommand{\vvvI}{\vvv_1}
\newcommand{\vvvIO}{\vvv_1^{(0)}}

\newcommand{\rhoO}{\rho_0}
\newcommand{\rhoOO}{\rho_0^{(0)}}

\newcommand{\rhoHat}{\hat{\rho}}

\newcommand{\SIum}{\upmu\textrm{m}}

\newcommand{\SIMHz}{\textrm{MHz}}

\newcommand{\SImm}{\textrm{mm}}

\newcommand{\SIs}{\textrm{s}}

\newcommand{\beq}[1]{\begin{equation} \eqlab{#1}}
\newcommand{\eeq}{\end{equation}}
\newcommand{\bsub}{\begin{subequations}}
\newcommand{\esub}{\end{subequations}}
\def\bal#1\eal{\begin{align}#1\end{align}}
\def\bsubal#1\esubal{\bsub \begin{align}#1\end{align} \esub}
\newcommand{\nn}{\nonumber}
\newcommand{\eqlab}[1]{\label{eq:#1}}
\renewcommand{\eqref}[1]{Eq.~(\ref{eq:#1})}

\newcommand{\eqrefnoEq}[1]{(\ref{eq:#1})}
\newcommand{\eqsref}[2]{Eqs.~(\ref{eq:#1}) and~(\ref{eq:#2})}

\newcommand{\figref}[1]{Fig.~\ref{fig:#1}}

\newcommand{\figlab}[1]{\label{fig:#1}}

\newcommand{\secref}[1]{Section~\ref{sec:#1}}

\newcommand{\seclab}[1]{\label{sec:#1}}

\newcommand{\sigmabf}{\bm{\sigma}}

\newcommand{\Rnm}{R_{nm}}
\newcommand{\Cnm}{C_{nm}}
\newcommand{\fnm}{f_{nm}}
\newcommand{\cOO}{c_0^{(0)}}
\newcommand{\cHat}{\hat{c}}
\begin{document}
%\preprint{Preprint identifier}

\title{Acoustic Tweezing and Patterning of Concentration Fields in Microfluidics}

\author{Jonas T. Karlsen}
\email{jonkar@fysik.dtu.dk}
\affiliation{Department of Physics, Technical University of Denmark, DTU Physics Building 309, DK-2800 Kongens Lyngby, Denmark}

%\author{Per Augustsson}
%\affiliation{Department of Biomedical Engineering, Lund University, Ole R\"{o}mers v\"{a}g 3, 22363, Lund, Sweden}

\author{Henrik Bruus}
\email{bruus@fysik.dtu.dk}
\affiliation{Department of Physics, Technical University of Denmark, DTU Physics Building 309, DK-2800 Kongens Lyngby, Denmark}

\date{5 December 2016}

\begin{abstract}
We demonstrate theoretically that acoustic forces acting on inhomogeneous fluids can be used to pattern and manipulate solute concentration fields into spatio-temporally controllable configurations stabilized against gravity. A theoretical framework describing the dynamics of concentration fields that weakly perturb the fluid density and speed of sound is presented and applied to study manipulation of concentration fields in rectangular-channel acoustic eigenmodes and in Bessel-function acoustic vortices. In the first example, methods to obtain horizontal and vertical multi-layer stratification of the concentration field at the end of a flow-through channel are presented. In the second example, we demonstrate acoustic tweezing and spatio-temporal manipulation of a local high-concentration region in a lower-concentration medium, thereby extending the realm of acoustic tweezing to include concentration fields.
\end{abstract}

% \pacs{43.25.Qp, 43.20.Fn, 43.20.+g, 47.35.Rs}

% 43 Acoustics
%   43.20.+g: General linear acoustics
%   43.20.Fn: Scattering of acoustic waves
%   43.20.Ks: Standing waves, resonance, normal modes
%	43.25.Nm: Acoustic streaming
%   43.25.Qp: Radiation pressure
%	43.80.-n, 43.80.+p: Ultrasound application to biology
% 47 Fluid dynamics
%   47.15.-x: laminar
%	47.35.Rs: Sound waves in fluids

%\keywords{Suggested keywords} Use showkeys class to display keywords

\maketitle

% Main text

\section{Introduction}

Sparked by the ambition to dynamically manipulate microparticles in solution, there have been major advances in the development of experimental methods to control ultrasound acoustic fields at the microscale~\cite{Drinkwater2016, Bruus2011c}, for example, using bulk acoustic waves~\cite{Laurell2007, Augustsson2011, Leibacher2014}, surface acoustic waves~\cite{Ding2012, Tran2012, Riaud2015a}, transducer arrays~\cite{Courtney2014, Marzo2015, Baresch2016}, and 3d-printed transmission holograms~\cite{Melde2016}. The acoustic radiation force acting on particles in acoustic fields is used in these systems to manipulate particles and cells, thereby concentrating~\cite{Antfolk2015}, trapping~\cite{Wiklund2014}, separating~\cite{Lee2015}, and sorting~\cite{Grenvall2015} bioparticles and cells based on their acoustomechanical properties. It would be of considerable interest if these methods could be extended to the manipulation of solute concentration fields in microfluidic systems. Indeed, the ability to pattern and manipulate molecular concentration fields plays essential roles in several lab-on-a-chip applications and in controlled studies of biological processes such as development, inflammation, wound healing, and cancer, for which biomolecule gradients act as cellular signaling mechanisms~\cite{Keenan2008}. The standard approach to precisely generate specified concentration gradients is to use microfluidic networks~\cite{Takayama1999, Dertinger2001}, however, with limited temporal control.

Here, we present a theoretical analysis of acoustic tweezing, patterning, and manipulation of solute concentration fields in microfluidics. We predict that acoustics offers a high degree of spatio-temporal control in these dynamical operations. Our study is predominantly motivated by the recent development of iso-acoustophoresis~\cite{Augustsson2016}, a microfluidic analog to density-gradient centrifugation. In iso-acoustophoresis cells are differentiated by their phenotype-specific acoustic impedance by observing their equilibrium position in an acoustically stabilized concentration gradient. The physics of this stabilization was only recently understood~\cite{Karlsen2016}, and an increased understanding of the ability of acoustics to shape and manipulate a concentration field is important to further develop the method.

In this work, we explore the consequences of our recent theory of the acoustic force density acting on inhomogeneous fluids in acoustic fields~\cite{Karlsen2016}, a theory successfully validated by experiments, which explains the acoustic stabilization and relocation of inhomogeneous fluids observed in microchannels~\cite{Deshmukh2014}. We define an inhomogeneous fluid as a fluid with spatial variations in density and speed of sound caused by a varying concentration of a solute. Consequently, there is a direct correspondence between fluid inhomogeneities and solute concentration. We present a theoretical framework for analyzing acoustic manipulation of such concentration fields, and apply it to the special cases of rectangular-channel eigenmodes and Bessel-function acoustic vortices. In the former system, we present methods to obtain stable horizontal and vertical multi-layer stratification of the concentration field at the end of a flow-through channel starting from typical inlet conditions. In the latter system, we demonstrate acoustic tweezing and spatio-temporal manipulation of a local high-concentration region in a lower-concentration medium. This extends the realm of acoustic tweezing to include concentration fields.

\begin{figure*}[!t]
\centering
\includegraphics[width=2.0\columnwidth]{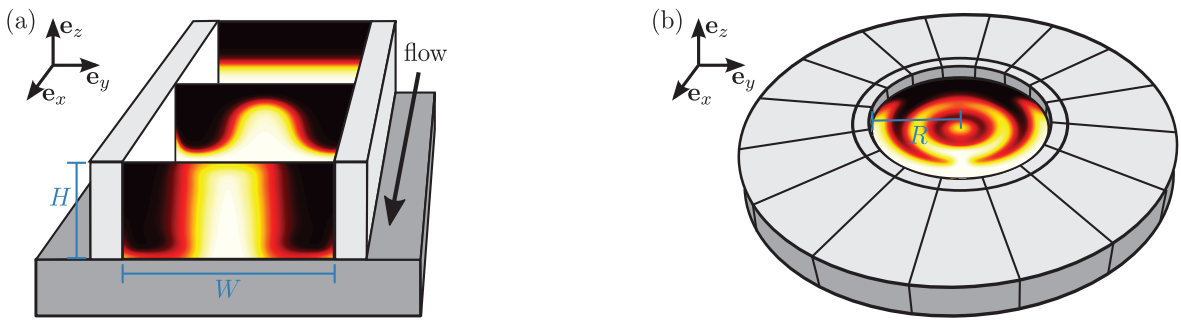}
\caption[]{\figlab{fig_01} (Color online)
Sketches of the two model systems considered in this work for the controlled ultrasound manipulation of inhomogeneous fluids at the microscale. The concentration fields (white, high; black, low) are manipulated by the acoustic field excited in the fluid domain by the attached piezoelectric transducers. (a) Acoustic eigenmodes in the two-dimensional cross-section of a rectangular microchannel of width $W$ and height $H$ in the $yz$-plane. (b) Acoustic Bessel-function vortices in the two-dimensional $xy$-plane generated by a circular 16-element phased transducer array of inner radius $R$. Gravity acts in the negative $z$-direction.}
\end{figure*}

\section{Model systems}
\seclab{model_systems}

In~\figref{fig_01}, the two typical model systems are introduced to provide the context necessary to appreciate the ensuing theoretical development. The implementation and design of the numerical model and how it corresponds to experimental conditions is discussed in more detail in~\secref{numerical_model}.

The first model system, shown in~\figref{fig_01}(a), is a long, straight, rectangular glass-silicon microchannel, placed along the $x$-axis, with a piezo-transducer glued underneath. By actuating the transducer at a resonance frequency of the cavity, an acoustic standing wave field can be established in the channel cross-section in the $yz$-plane, which is typically a few hundred $\SIum$ in the width $W$ and height $H$ leading to fundamental resonance frequencies of order 1-10$~\SIMHz$. These systems are well-characterized~\cite{Barnkob2010, Augustsson2011, Muller2013, Oever2015, Lamprecht2016} and used in various biomedical applications, for example, the enrichment of circulating tumor cells in blood~\cite{Augustsson2012, Antfolk2015}.

The second model system, shown in~\figref{fig_01}(b), consists of a transducer array with 16 elements enclosing a circular fluid chamber. It is inspired by, and closely resembles, the experimental systems in Refs.~\cite{Bernassau2013, Courtney2013, Courtney2014, Riaud2015}. The radius of these chambers is typically around 1~mm, and the chambers may have between 8 and 64 transducer elements operating at MHz frequency. By controlling the amplitude and phase of each transducer, approximate Bessel-function acoustic vortices may be generated by a superposition of waves, and then used to trap and move microparticles~\cite{Courtney2013, Courtney2014}.

\section{Theory}
\seclab{theory}

The recently developed theory for the acoustic force density acting on inhomogeneous fluids in acoustic fields~\cite{Karlsen2016} is based on the separation of time scales between the fast acoustic time scale $t$ and the slow hydrodynamic time scale $\tau$. In general, the large separation of time scales ($\tau\sim10^5t$) allows the acoustic fields, oscillating at the fast time scale $t$, to be solved for while keeping the hydrodynamic degrees of freedom fixed at each instance in time $\tau$ on the slow time scale. Due to the inhomogeneity in the fluid medium, the resulting acoustic field yields a divergence in the time-averaged acoustic momentum-flux-density tensor~\cite{Karlsen2016}, and this is the origin of the acoustic force density $\fffac$, which enters the slow-time-scale hydrodynamics as an external driving force.

The inhomogeneity in the fluid medium is caused by the solute concentration field $s(\rrr,\tau)$. The fluid density $\rhoO$, compressibility $\kapO$, and dynamic viscosity $\etaO$ are all functions of the solute concentration $s$, and thus functions of space and time as the concentration field evolves by advection and diffusion,
 \beq{inhoms}
 \rhoO=\rhoO\big(s(\rrr,\tau)\big) , \, \kapO=\kapO\big(s(\rrr,\tau)\big) , \, \etaO=\etaO\big(s(\rrr,\tau)\big) .
 \eeq
The specific dependence of $\rhoO$, $\kapO$, and $\etaO$ on concentration $s$ depend on the solute used to establish the inhomogeneity, e.g. iodixanol (OptiPrep), polysucrose (Ficoll), or colloidal nanoparticles (Percoll) as commonly used in density-gradient centrifugation. In this work we consider solutions of iodixanol, for which we have measured the fluid properties as functions of concentration~\cite{Augustsson2016}.

\subsection{Slow-time-scale hydrodynamics}

The hydrodynamics on the slow time scale $\tau$ is governed by the momentum- and mass-continuity equations for the fluid velocity $\vvv(\rrr,\tau)$ and pressure $p(\rrr,\tau)$, as well as the advection-diffusion equation for the solute concentration field $s(\rrr,\tau)$ of the solute with diffusivity $D$,
 \bsub
 \eqlab{DynamicsSlow}
 \bal
 \eqlab{NSSlow}
 \pp_\tau (\rhoO \vvv) &= \divop \big[ \sigmabf - \rhoO\vvv\vvv \big] + \fffac + \rhoO \gvec , \\
 \eqlab{ContSlow}
 \pp_\tau \rhoO &= - \divop \big( \rhoO \vvv \big) , \\
 \eqlab{DiffusionSlow}
 \pp_\tau s &= - \divop \big[ - D \nablabf s + \vvv s \big] .
 \eal
 \esub
Here, $\gvec$ is the acceleration due to gravity, and $\sigmabf$ is the fluid stress tensor, given by
 \bal
 \sigmabf = - p \, \mathbf{I} + \etaO \Big[ \nablabf \vvv + (\nablabf \vvv)^\mathrm{T} \Big] + \Big(\etaO^\mathrm{b} -\frac{2}{3}\etaO \Big) (\divop\vvv) \, \mathbf{I} ,
 \eal
where the superscript $\mathrm{T}$ indicates tensor transposition, and $\etaO$ and $\etaO^\mathrm{b}$ are the dynamic and bulk viscosity, respectively. The equations constitute an advection-diffusion flow problem with an external forcing due to the acoustic and gravitational force densities $\fffac$ and $\rhoO \gvec$, both appearing on the right-hand side of the momentum equation~\eqrefnoEq{NSSlow}.

\subsection{The acoustic force density}

The acoustic force density $\fffac$ acting on the fluid on the slow hydrodynamic time scale $\tau$ was derived in Ref.~\cite{Karlsen2016} from a divergence in the time-averaged acoustic momentum-flux-density tensor induced by continuous spatial variations in the fluid parameters of density $\rhoO$ and compressibility $\kapO$,
 \beq{facFinal}
 \fffac = - \frac14 |\pI|^2 \nablabf\kapO - \frac14 |\vvvI|^2 \nablabf\rhoO .
 \eeq
Here, $\pI$ and $\vvvI$ are the acoustic pressure and velocity field, respectively, assumed to be time-harmonic first-order perturbations of the hydrodynamic degrees of freedom.

Because the compressibility $\kapO$ is difficult to measure directly, it is often more convenient to work with the fluid density $\rhoO$ and speed of sound $c_0$, both of which are readily measured as functions of concentration. Using that $\kapO=1/(\rhoO c_0^2)$, we find
 \beq{dkapO}
 \nablabf \kapO = \nablabf \Big( \frac{1}{\rhoO c_0^2} \Big)
 = - \kapO \frac{\nablabf \rhoO}{\rhoO} - 2\kapO\frac{\nablabf c_0}{c_0} ,
 \eeq
and the expression~\eqrefnoEq{facFinal} becomes
 \beq{fac_crho}
 \fffac = \frac{1}{4} \Big[ \kapO |p_1|^2 - \rhoO |\vvvI|^2 \Big] \frac{\nablabf \rhoO}{\rhoO}
 + \frac12 \kapO |p_1|^2 \frac{\nablabf c_0}{c_0} .
 \eeq
In the weakly inhomogeneous limit where the variations in density $\rhoO$ and speed of sound $c_0$ are small, we introduce the dimensionless relative deviations $\rhoHat(\rrr,\tau)$ and $\cHat(\rrr,\tau)$, and write
 \bsub
 \eqlab{weakInhom}
 \bal
 \rhoO(\rrr,\tau) = \rhoOO[1 + \rhoHat(\rrr,\tau)] , \quad |\rhoHat(\rrr,\tau)| \ll 1 , \\
 c_0(\rrr,\tau) = \cOO[1 + \cHat(\rrr,\tau)] , \quad |\cHat(\rrr,\tau)| \ll 1 .
 \eal
 \esub
Here, the superscript $(0)$ indicates zeroth-order in the inhomogeneity $\rhoHat$ and $\cHat$. To first order in $\rhoHat$ and $\cHat$ the force density~\eqrefnoEq{fac_crho} then becomes
 \bal
 \eqlab{facWeakInhom}
 \fffacI = \frac14 \Big[ \kapOO |\pIO|^2 - \rhoOO |\vvvIO|^2 \Big] \nablabf\rhoHat
 + \frac12 \kapOO |\pIO|^2 \nablabf\cHat .
 \eal
In this expression, the acoustic fields $\pIO$ and $\vvvIO$ are zeroth order in $\rhoHat$ and $\cHat$, and consequently, the fields are obtained as solutions of the homogeneous-fluid wave equation. This constitutes a significant simplification in applications of the theory, as will be shown next.

Let $\pa$ denote the acoustic pressure amplitude, $\omega$ the angular acoustic frequency, and $\kOO=\omega/\cOO$ the homogeneous-fluid wave number. The time-harmonic acoustic fields $\pIO$ and $\vvvIO$ may then be written in terms of a non-dimensionalized pressure field $\pIOh(\rrr,\tau)$, as
 \beq{fieldsGeneral}
 \pIO = \pa \pIOh \ee^{-\ii\omega t}, \quad
 \vvvIO = \frac{ - \ii \pa}{\kOO \cOO \rhoOO} \nablabf \pIOh \ee^{-\ii\omega t}.
 \eeq
Inserting this into~\eqref{facWeakInhom} and introducing the homogeneous-fluid oscillation-time-averaged acoustic energy density $\EacO=\frac{1}{4}\kapOO\pa^2$, the acoustic force density $\fffacI$ can be rewritten as
 \bsub
 \eqlab{facRC}
 \bal
 \eqlab{facRCa}
 \fffacI = \EacO \Big[ R(\rrr,\tau)\: \nablabf \rhoHat + C(\rrr,\tau)\: \nablabf \cHat \Big] ,
 \eal
where we have introduced the dimensionless field-shape functions $R(\rrr,\tau)$ and $C(\rrr,\tau)$, given by
 \bal
 R(\rrr,\tau) &= |\pIOh|^2 - \big(\kOO\big)^{-2} |\nablabf \pIOh|^2 , \\
 C(\rrr,\tau) &= 2 |\pIOh|^2 .
 \eal
 \esub
The field-shape functions $R(\rrr,\tau)$ and $C(\rrr,\tau)$ depend on the shape of the homogeneous-fluid acoustic pressure field $\pIOh(\rrr,\tau)$, often known analytically, and may thus be varied in space and time. Consequently, our theoretical framework suggests that a high level of spatio-temporal control of fluid inhomogeneities can be achieved.

\subsection{Eigenmodes in a rectangular microchannel}

Consider a long, straight, hard-walled microchannel of width $W$ and height $H$, with the aspect ratio $\alpha=H/W$. The acoustic fields obtained at resonance conditions in the two-dimensional channel cross-section take the form of eigenmode solutions to the Helmholtz wave equation with hard-wall boundary conditions. Choosing the fluid domain in the $yz$-plane defined by $0<y<W$ and $0<z<H$, and introducing the normalized coordinates $\yhat=\frac{\pi}{W} y$ and $\zhat = \frac{\pi}{H} z$, the eigenmodes $\pIOh(\yhat,\zhat)$ are
 \bsub
 \eqlab{fieldsRect}
 \bal
 \pIOh &= \cos( n \yhat) \cos( m \zhat) , \\
 \eqlab{fres_nm}
 \mathrm{with} \quad \fnm &= \frac{\omega_{nm}}{2\pi} = \frac{c}{2} \sqrt{ \Big( \frac{n}{W} \Big)^2 + \Big( \frac{m}{H} \Big)^2}.
 \eal
 \esub
Here, $n=0,1,2,...$ and $m=0,1,2,...$ are the mode numbers in the $y$- and $z$-direction, respectively, and $\fnm$ is the resonance frequency of the $nm$-mode.

Inserting the eigenmode solution~\eqrefnoEq{fieldsRect} into~\eqref{facRC}, one obtains the acoustic force density acting on the fluid in the $nm$-mode. After some algebra, the field-shape functions $\Rnm(\yhat,\zhat)$ and $\Cnm(\yhat,\zhat)$ take the form,
 \bsub
 \eqlab{RCeigen}
 \bal
 \Rnm(\yhat,\zhat) &= \frac12 \Bigg\lbrace
 \frac{n^2}{n^2 + m^2 \alpha^{-2}} \Big[ \cos(2n\yhat) - \cos(2m\zhat) \Big] \nn \\
 & \quad\quad + \cos(2n\yhat)\cos(2m\zhat) + \cos(2m\zhat)  \Bigg\rbrace , \\
 \Cnm(\yhat,\zhat) &= \frac12 \Big[1 + \cos(2n\yhat) \Big] \Big[1 + \cos(2m\zhat) \Big] .
 \eal
 \esub
In the horizontal half-wave resonance $(n,m)=(1,0)$, we obtain $R_{10}=\cos(2\yhat)$ and $C_{10}=1+\cos(2\yhat)$, in agreement with Ref.~\cite{Karlsen2016}, given an appropriate change of the coordinate system.

\subsection{Bessel-function acoustic vortex fields}

It has been demonstrated that transducer arrays can be used to generate acoustic vortices in fluid-filled chambers~\cite{Courtney2013, Courtney2014, Riaud2015, Hefner1999}. By controlling the amplitude and phase of each transducer in a circular array, one can generate approximate Bessel-function pressure fields of the form~\cite{Courtney2013},
 \beq{fieldsBessel}
 \pIOh = J_l(\kOO r) \ee^{\ii l \theta} .
 \eeq
Here, we are using cylindrical polar coordinates $(r,\theta,z)$ with the origin at the center of the Bessel function. $J_l$ is the $l$'th order Bessel function of the first kind, and $l$ is the number of $2\pi$ phase shifts around the axis of the vortex, often referred to as the topological charge.

The acoustic force density acting on an inhomogeneous fluid in the acoustic vortex is obtained by inserting~\eqref{fieldsBessel} into~\eqref{facRC}. Introducing the normalized radial coordinate $\rhat=\kOO r$, and making use of the recurrence relations $\frac{2 n}{\rhat} J_n(\rhat) = J_{n-1}(\rhat) + J_{n+1}(\rhat)$ and $2 J_n^\prime(\rhat) = J_{n-1}(\rhat)-J_{n+1}(\rhat)$, the field-shape functions $R_l(\rhat)$ and $C_l(\rhat)$ of the $l$'th order vortex take the form,
 \bsub
 \eqlab{RCbessel}
 \bal
 \eqlab{RCbesselA}
 R_l(\rhat) &= [J_l(\rhat)]^2 - \frac12 [J_{l-1}(\rhat)]^2 - \frac12 [J_{l+1}(\rhat)]^2 , \\
 C_l(\rhat) &= 2 [J_l(\rhat)]^2 .
 \eal
 \esub

\section{Numerical model}
\seclab{numerical_model}

In this section we present the implementation and design of our numerical models. Emphasis is put on the considerations that went into designing numerical models that describe actual experimental conditions that may be reproduced with the setups introduced in~\secref{model_systems} and sketched in \figref{fig_01}.

\subsection{Numerical implementation}

In the numerical models of the slow-time-scale hydrodynamics, the coupled field equations~\eqrefnoEq{DynamicsSlow} are implemented and solved on weak form using the finite-element solver COMSOL Multiphysics~\cite{COMSOL52}. We consider the limit of weakly inhomogeneous fluids and use the analytical expression~\eqrefnoEq{facRCa} for the acoustic force density $\fffacI$ with the field-shape functions given in the rectangular-channel eigenmodes and acoustic vortex fields, respectively, in~\eqsref{RCeigen}{RCbessel}. For numerical stability, a logarithmic concentration field $\hat{s}$, with $s=s_0 \exp(\hat{s})$, is used as the independent concentration variable.

The boundary conditions imposed on the slow-time-scale velocity and concentration fields $\vvv(\rrr,\tau)$ and $s(\rrr,\tau)$ at the boundary $\pp\Omega$ of the fluid domain $\Omega$ with normal vector $\nnn$, are the standard no-slip and no-flux conditions,
 \bal
 \eqlab{BCs}
 \vvv=\vec{0}, \quad \nnn\cdot\nablabf s = 0 , \quad \mathrm{for} \ \rrr\in\pp\Omega ,
 \eal
Several convergence tests were carried out to ensure numerical convergence. For example, the integrated concentration was conserved with a maximum relative error of $2\times10^{-3}$ at all times.

\subsection{Modeling the fluid inhomogeneity}

We model aqueous solutions of iodixanol (OptiPrep), for which the fluid parameters have been measured experimentally as functions of the iodixanol volume-fraction concentration $s$~\cite{Augustsson2016}. OptiPrep is a cell-friendly medium that is used in density-gradient centrifugation and iso-acoustic focusing. In the models, we consider initial concentration fields with iodixanol volume-fractions ranging from $s_\mathrm{min}=0.1$ to $s_\mathrm{max}=0.3$, yielding a relative density difference of up to 10\%, while the maximum relative variation in the speed of sound is 0.5\%. Consequently, we neglect variations in $\cO$, which means that only gradients in $\rhoO$ contribute to the acoustic force density.

The polynomials fitting the measured density $\rhoO(s)$ and dynamic viscosity $\etaO(s)$, as functions of the iodixanol volume-fraction concentration $s$, are~\cite{Augustsson2016}
 \bsubal
 \rhoO &= \rhoOO[1+ a_1 s] ,  \\
 \etaO &= \etaOO[1+b_1 s + b_2 s^2 + b_3 s^3] .
 \esubal
Here, $\rhoOO = 1005~\mathrm{kg/m}^3$ and $\etaOO = 0.954~\mathrm{mPa\,s}$, and the dimensionless constants are $a_1=0.522$, $b_1=2.05$, $b_2=2.54$, and $b_3=22.8$. The diffusivity of iodixanol was measured to $D=0.9\times10^{-10}$~$\mathrm{m^2/s}$. For the bulk viscosity we use the value of pure water~\cite{Muller2014}.

\begin{figure*}[!t]
\centering
\includegraphics[width=2.0\columnwidth]{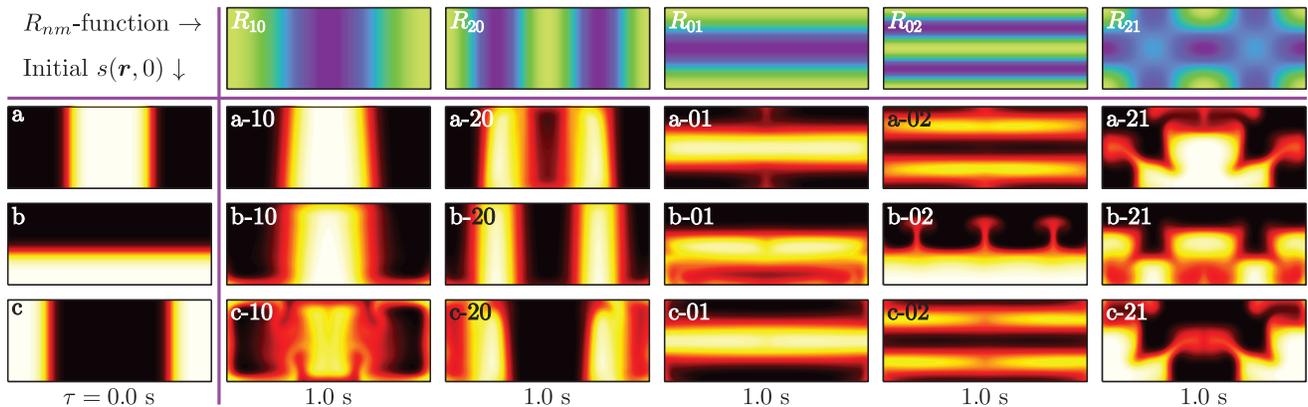}
\caption[]{\figlab{fig_02} (Color online)
Patterning of inhomogeneous iodixanol solutions in rectangular-channel eigenmodes. The top row shows the field-shape functions $R_{nm}$ for each mode $nm$ (min, dark blue; max, light green). Three different initial concentration fields $s(\rrr,0)$ of the dense (30\% iodixanol, white) and less dense (10\% iodixanol, black) solutions are considered (1st column, $a$, $b$ and $c$). The next columns show the resulting concentration fields $s(\rrr,\tau)$ after a retention time of $\tau=1.0~\SIs$ in either the 10-mode (2nd column), the 20-mode (3rd column), the 01-mode (4th column), the 02-mode (5th column), or the 21-mode (6th column), starting from the initial condition $a$ (second row), $b$ (third row), or $c$ (bottom row).}
\end{figure*}

\subsection{Modeling the rectangular microchannel}

In this model, we consider a long straight rectangular microchannel of width $W=375~\SIum$ and height $H=150~\SIum$ as sketched in~\figref{fig_01}(a). In acoustophoresis experiments, acoustic eigenmodes of the two-dimensional channel cross-section transverse to the flow are used extensively to manipulate and focus particles and cells based on their mechanical properties. Two notable advantages of using acoustic eigenmodes, or bulk acoustic waves, is that the eigenmodes are easily excited by an attached piezoceramic transducer actuated at the resonance frequency, and that high acoustic energy densities can be obtained in the resonant modes. Typical quality-factors in glass-silicon microchips are between $10^2$ and $10^3$, and typical measured acoustic energy densities are in the range 1-1000~J/m$^3$~\cite{Barnkob2010, Oever2015}. We use $\EacO=10~\mathrm{J/m^3}$, approximately an order of magnitude larger than the hydrostatic pressure difference across the channel height, ensuring that gravity plays only a minor role in the fluid relocation~\cite{Karlsen2016}.

Referring again to \figref{fig_01}(a), we are modeling a flow-through microchannel system where the flow-rate can be controlled, thereby setting the retention time of the fluid in the channel. In our time-dependent model, the time $\tau$ can thus be translated into a downstream length $L$ from the inlet. For example, in the system under consideration a fluid retention time of $\tau_\mathrm{ret}=1.0$~s over a length of $L=5.0$~mm implies a flow-rate of $17~\upmu\mathrm{L/min}$, all of which are realistic experimental parameters. Diffusion generally plays an important role in manipulating concentration fields. However, the time scale of diffusion across one third of the channel width is $\tau_\mathrm{diff}=\frac{1}{2D}(\frac{1}{3}W)^2=87$~s, leaving enough time to conduct typical steady-flow experiments at relevant flow rates without diffusion flattening the gradients.

\subsection{Modeling the acoustic vortex field}

In this model, we consider a circular fluid chamber, as sketched in~\figref{fig_01}(b), in which an acoustic vortex field of the form~\eqrefnoEq{fieldsBessel} is excited by the surrounding transducer array or by swirling surface acoustic waves~\cite{Courtney2014, Riaud2015}. Notice that, in contrast to the rectangular-microchannel acoustic fields, the acoustic vortices are non-resonant fields, and the center of the vortex can be moved relative to the chamber. In our model, we use a chamber of radius $R=250~\SIum$, an acoustic energy density of $\EacO=10~\mathrm{J/m^3}$, and a frequency of $f = 7.5~\SIMHz$.

\section{Simulation results}
We present a selection of simulation results demonstrating acoustics as a means to spatio-temporally control, manipulate, and relocate solute concentration fields in microsystems. Specifically, we demonstrate manipulation of concentration fields in rectangular-channel eigenmodes and in acoustic vortex fields in circular chambers. In the former, we demonstrate the use of sequential eigenmode actuation to obtain horizontal or vertical multi-layering of the fluid inhomogeneities. We further motivate and introduce the simple but useful concept of orthogonal relocation. In the circular chamber, we demonstrate trapping and translation of a fluid inhomogeneity using Bessel-function acoustic tweezers.

\subsection{Multi-layering of concentration fields in rectangular-channel eigenmodes}

We consider patterning of concentration fields in the $nm$-eigenmodes in the rectangular microchannel using the modes $(n,m)=(1,0)$, $(2,0)$, $(0,1)$, $(0,2)$, and $(2,1)$ as examples. The resonance frequency $\fnm$ of these eigenmodes is obtained from~\eqref{fres_nm}, yielding $f_{10} = 2.0~\SIMHz$, $f_{20} = 4.0~\SIMHz$, $f_{01} = 5.0~\SIMHz$, $f_{02} = 10~\SIMHz$, and $f_{21} = 6.4~\SIMHz$.

In~\figref{fig_02} we consider three different initial conditions $a$, $b$, and $c$ (first column) on the concentration field $s(\rrr,0)$. In the following columns are shown the concentration fields $s(\rrr,\tau)$ in the selected $nm$-modes after a time $\tau=1.0~\SIs$, for each of the three initial configurations $a$, $b$, and $c$. The resulting configurations are denoted $i$-$nm$, with $i$ indicating the initial configuration ($i=a$, $b$, or $c$), and $nm$ denoting the mode of actuation. The top row shows the field-shape functions $R_{nm}$ of the corresponding modes. In general, the denser high-concentration fluid (30\% iodixanol, white) is relocated into the minima of $R_{nm}$ appearing at pressure nodes, as one might anticipate from the analogy to the acoustic radiation force acting on a particle. It should be emphasized, however, that in contrast to the acoustic radiation force acting on a particle in a standing wave, $\fffac$ is a non-conservative force and it cannot in general be written as the gradient of a potential. The acoustic force density $\fffac$ moreover depends on the history of the system, which is also in contrast to the particle force. For a given mode, the concentration fields tend to evolve towards the same quasi-stable equilibrium configuration, however, the different initial conditions generally influence the resulting configurations.

Inspecting~\figref{fig_02}, one finds that relocation of the inhomogeneity into vertical layers is obtained for $m=0$, while horizontal layers are obtained for $n=0$. This is to be expected from the geometry of the acoustic field. However, comparing $a$-01, $b$-01, and $c$-01 it is evident that the concentration field after 1 s of actuation in the 01-mode depends strongly on the initial configuration ($a$, $b$, or $c$). Indeed, the configurations $a$ and $c$ have been relocated into much "cleaner" 01-mode configurations with a single horizontal layer as compared to the configuration $b$. The reason is that the relocations $a \rightarrow a$-01 and $c \rightarrow c$-01 are orthogonal relocations in the sense that the initial and final stratifications are orthogonal to one another. In contrast, the relocation $b \rightarrow b$-01 is a parallel relocation, where whole fluid layers are to be moved into new parallel positions, which can only proceed by an instability. This is particularly evident in the 02-mode comparing the orthogonally relocated configurations $a$-02 and $c$-02 to $b$-02, the latter for which the parallel relocation proceeds by a Rayleigh--Taylor-like instability, shooting up three streams that slowly feed the second horizontal layer.

\begin{figure}[!t]
\centering
\includegraphics[width=0.92\columnwidth]{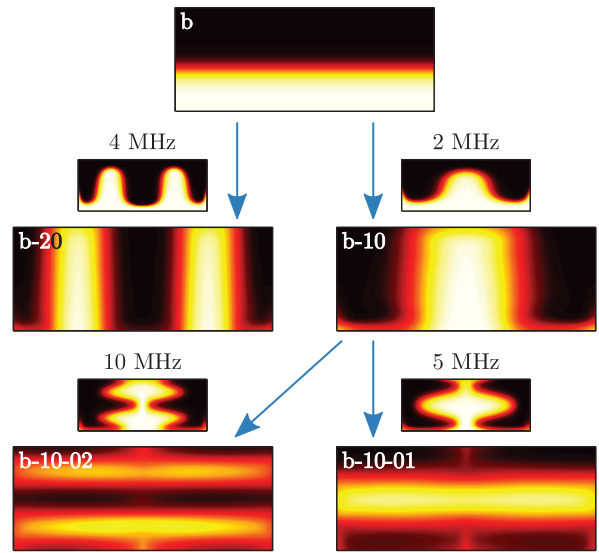}
\caption[]{\figlab{fig_03} (Color online)
Vertical and horizontal layering of iodixanol concentration fields $s(\rrr,\tau)$ starting from the horizontally-layered initial configuration $b$ with the dense fluid (30\% iodixanol, white) at the bottom of the channel and the less dense fluid (10\% iodixanol, black) at the top (top row). Each arrow (blue) represents an orthogonal relocation obtained by exciting an $nm$-eigenmode in the rectangular channel for 1 s, with the miniature showing the transition 50~ms after the mode shift. Vertical layering is obtained directly by actuation of the 10- or the 20-mode, yielding the configurations $b$-10 and $b$-20, respectively. Horizontal layering involves an intermediate step going through the 10-mode, yielding the $b$-10-01 and $b$-10-02 configurations. }
\end{figure}

These observations suggest that orthogonal relocation provides the most effective way of relocating and patterning concentration fields. In the event that a desired relocation is parallel, as in the example $b$-01 starting from the configuration $b$, the resulting horizontally layered 01-mode configuration is blurred because it proceeded by an instability. The solution to obtaining sharp horizontally-layered 01- and 02-mode configurations starting from $b$ is to go through a sequence of orthogonal relocations. By applying the sequence $b$-10-0$m$, the 10-mode being an intermediate, one can achieve sharp horizontally-layered 0$m$-mode configurations from the initial configuration $b$. This is illustrated in~\figref{fig_03}, where the relocation dynamics is also indicated by showing intermediate configurations. A movie of the dynamics in the sequence $b$-10-01-20 can be found in the Supplemental Material~\footnote{See Supplemental Material at [url] for movies of the time-evolution of the concentration fields.}.

\begin{figure*}[!t]
\centering
\includegraphics[width=2.0\columnwidth]{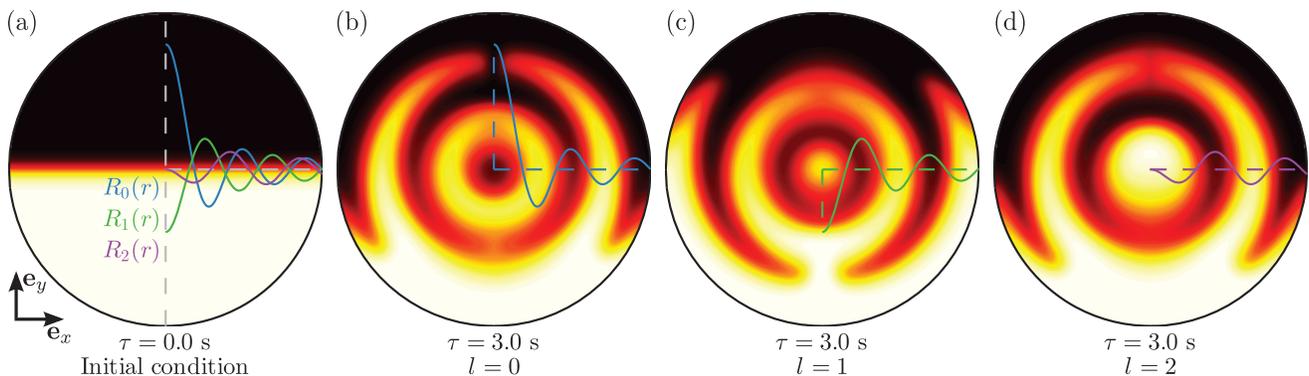}
\caption[]{\figlab{fig_04} (Color online)
Patterning of inhomogeneous iodixanol solutions in acoustic vortices of topological charge $l$. (a) Initial concentration field $s(\rrr,0)$ with the dense (30\% iodixanol, white) and less dense (10\% iodixanol, black) solution each occupying half of the circular domain. The radial field-shape function $R_l(r)$ is shown for $l=0$ (blue), $l=1$ (green), and $l=2$ (violet), indicating the initial magnitude and (negative) direction of the acoustic force density acting on the blurred interface. (b)-(d) Resulting concentration fields $s(\rrr,\tau)$ after $\tau=3.0$~s in the acoustic vortex with $l=0$, $l=1$, and $l=2$, respectively, with a central trapping region for $l>0$. The denser fluid (white) is relocated into the minima of the field-shape functions $R_l$.}
\end{figure*}

In summary, starting from a single-layer configuration, one can achieve multi-layering of concentration fields on a one-second timescale in the rectangular-channel eigenmodes commonly employed in acoustophoresis. While we have focused on the spatial patterning, the ability to switch between modes provides temporal control of the concentration field at the end of the flow-through channel. This type of acoustic fluid manipulation is best performed by orthogonal relocation, and a parallel relocation can always be substituted by two sequential orthogonal relocations.

\subsection{Patterning and tweezing of concentration fields in acoustic vortex fields}

Next, we demonstrate patterning and spatio-temporal manipulation of concentration fields in Bessel-function acoustic vortex fields in circular fluid chambers. Starting from the initial concentration field $s(\rrr,0)$, shown in~\figref{fig_04}(a), with the denser fluid (30\% iodixanol, white) occupying half the circular domain, \figref{fig_04}(b)-(d) shows the concentration fields $s(\rrr,\tau)$ after $\tau=3.0~\SIs$ of actuation in an acoustic vortex of order $l=0$, $l=1$, and $l=2$, respectively. Again, it is observed that the denser fluid tends to be relocated into the minima of the field-shape functions $R_l$.

The central region of an acoustic vortex is of particular interest because it provides a trapping potential that can be used to trap and manipulate particles. Here, considering inhomogeneous fluid manipulation, we define the central region of the $l$'th order vortex from the condition $\rhat < \rhat_l^*$, where $\rhat_l^*$ is the first non-zero root of the field-shape function, $R_l(\rhat_l^*)=0$. This yields the approximate values, $\rhat_0^*=1.44$, $\rhat_1^*=1.18$, and $\rhat_2^*=2.26$. As demonstrated in \figref{fig_04}, in the vortex with $l=0$ the denser fluid (white) is forced outside of the central region, while in the vortices with $l=1$ and $l=2$ the denser fluid is forced into the central region. Mathematically, this follows directly from \eqref{facRCa} (with $\nablabf\cHat=\vec{0}$) by inspecting the field-shape functions $R_l(r)$ shown in \figref{fig_04}, because they indicate the initial radial distribution of the acoustic force density acting on the blurred interface. Physically, the acoustic pressure is maximum at the center for $l=0$, while it is zero for $l>0$. Note furthermore, that for $l>0$ the central trapping region becomes larger for increasing $l$. These findings for manipulation of inhomogeneous fluids are analogous to those of acoustic tweezing of particles~\cite{Courtney2014}.

Acoustic tweezing of a high-concentration region in a lower-concentration medium can thus be realized in the central region of vortices with $l>0$, and this may be used to confine and translate a fluid inhomogeneity as will be demonstrated next using the $l=1$ vortex. We consider an initial concentration field $s(\rrr,0)$ that has a Gaussian high-concentration region (30\% iodixanol, white) centered at the position $(r,\theta)=(\frac12 R,\frac12 \pi)$, as given in polar coordinates, in the lower-concentration medium (10\% iodixanol), see \figref{fig_05}(a). The width (or standard deviation) of the Gaussian is set to $\sigma=0.5 \, \rhat_1^*$, half the width of the central trapping region. The acoustic vortex is initially centered at the position of the inhomogeneity, and it is then translated in a closed-loop equilateral triangle moving in straight lines from $(\frac12 R, \frac12 \pi)$ to $(\frac12 R,-\frac{1}{6}\pi)$, to $(\frac12 R,-\frac{5}{6}\pi)$, and finally back to the starting position in $(\frac12 R,\frac12 \pi)$. The translation speed $U=0.7~\SImm/\SIs$ of the center of the vortex was chosen such that it takes 0.3~s to move the distance from one corner of the triangle to the next. The resulting concentration field $s(\rrr,\tau)$ after $\tau=0.3$~s, 0.6~s, and 0.9~s is shown in~\figref{fig_05}(b), (c), and (d), respectively, with the central region of the vortex indicated by the green circle, and the path of the center of the vortex by the straight green lines. To a good approximation, the high-concentration solution is kept within the central region of the vortex as it is translated in space, leaving only a trailing diffusive residue. Movies showing the manipulation in real time for two different translation speeds are available in the Supplemental Material~\cite{Note1}. We find that when the translation speed of the vortex is increased by a factor of 3, the inhomogeneity does not remain trapped at the center during the full loop. Conversely, for slower translation speeds, the inhomogeneity stays in the center of the vortex, but the increased loop time leads to a more pronounced diffusion broadening.

\begin{figure*}[!t]
\centering
\includegraphics[width=2.0\columnwidth]{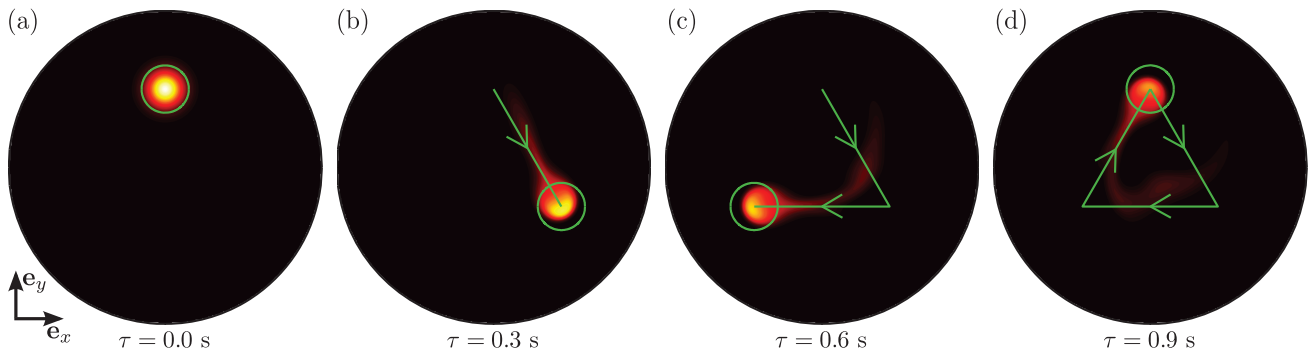}
\caption[]{\figlab{fig_05} (Color online)
Acoustic tweezing and translation of a local high-concentration region using an acoustic vortex with topological charge $l=1$. (a) Initial concentration field $s(\rrr,0)$ with a Gaussian high-concentration region (30\% iodixanol, white) in a lower-concentration medium (10\% iodixanol, black). Initially, the acoustic vortex is centered at the position of the inhomogeneity, with the green circle indicating the central region of the vortex. At time $\tau>0.0~\SIs$ the vortex is moved at constant speed $U=0.7~\SImm/\SIs$ along the green path in a closed-loop triangle. The resulting concentration fields $s(\rrr,\tau)$ after $\tau=0.3$~s, 0.6~s, and 0.9~s are shown in (b), (c), and (d), respectively.}
\end{figure*}

The results presented in this section provide theoretical evidence that the applicability of acoustic tweezers can be extended beyond particle manipulation to include manipulation of concentration fields -- a phenomenon that has yet to be demonstrated experimentally.

\section{Discussion}

In this paper, we have explored some consequences of our recent theory of the acoustic force density acting on inhomogeneous fluids~\cite{Karlsen2016}. For this purpose, a useful formulation of the theory was given in terms of the field-shape functions $R$ and $C$ in the experimentally relevant limit of weakly inhomogeneous fluids. The theory of the acoustic force density acting on inhomogeneous fluids show resemblance to the Gorkov theory of the acoustic radiation force acting on a particle~\cite{Gorkov1962}, for example, by the tendency of dense fluids being focused at the pressure nodes. However, the two theories have important distinctions. (1) The theory of the acoustic force density acting on inhomogeneous fluids is a field theory with $\fffac$ generally acting on the fluid in every point in space, in contrast to the Newtonian theory for the radiation force acting on a point particle. (2) The acoustic force density $\fffac$ is a non-conservative force, and in general it cannot be written as the gradient of a potential, as can the radiation force on a particle in a standing wave~\cite{Settnes2012, Karlsen2015}. Instead, one may use the field-shape functions to assess the direction and magnitude of the forces acting on the fluid for a given initial concentration field. For density inhomogeneities, the denser fluid tends to relocate to the minima of the field-shape function $R$. (3) Not unrelated, in the theory of the acoustic force density, the force density $\fffac$ depends on the history of the system and it evolves as the concentration field changes by advection and diffusion.

While the acoustic force density can stabilize a fluid inhomogeneity against destabilizing forces, such as gravity in the case of a density gradient, it cannot counteract molecular diffusion. Consequently an inhomogeneity always has a finite lifetime set by the characteristic diffusion time, and it will broaden due to diffusion. Interestingly, this is an advantage in iso-acoustic focusing, because it allows fine-tuning the gradient at the end of a steady-flow-through channel by varying the flow rate~\cite{Augustsson2016}. In acoustic tweezing of a high-concentration region, diffusion limits the time that the inhomogeneity can be manipulated in a closed chamber. One can obtain longer diffusion times by going to larger scales or by using Ficoll or Percoll solutions with larger solute molecules that diffuse slower.

Importantly, the ability to manipulate concentration fields requires that the concentration field introduces inhomogeneities ($\gtrsim 1\%$) in the fluid density or speed of sound. This is true for concentrations of iodixanol (OptiPrep), polysucrose (Ficoll), or colloidal nanoparticles (Percoll), that are used in density-gradient separation. To manipulate a concentration field of a specific biomolecule at low concentration, one can add OptiPrep, Ficoll, or Percoll, so the solution containing the dilute concentration of biomolecules still introduces a gradient.\\[10mm]

\section{Conclusion}

Advances in the development of experimental methods to control acoustic fields for microparticle-manipulation purposes, for example, using transducer arrays, surface acoustic waves, and transmission holograms, allows spatio-temporal tailoring of acoustic fields. In this paper, we have demonstrated theoretically that this provides dynamic control of solute concentration fields at the microscale. We can think of this as acoustic "landscaping" of concentration fields, because of the ability to dynamically manipulate "hills" and "valleys" of high and low concentration. Using acoustic landscaping one may relocate, shape, and pattern concentration fields with the methods already developed for particle-handling. We have presented two examples of this. Firstly, in rectangular microchannels, we have described an operational principle for obtaining multi-layer stratification of concentration fields using acoustic eigenmodes. Secondly, we have demonstrated acoustic tweezing and manipulation of a high-concentration fluid region in a lower-concentration fluid medium using a Bessel-function acoustic vortex.

We envision that the insights obtained in this study will find applications in the further development of iso-acoustophoresis and other gradient-based separation methods. Another use may be found in studies of biological processes with active spatio-temporal control of solute gradients. Finally, the ability to pattern fluid inhomogeneities using acoustics might also find applications in drug delivery, tissue engineering, and 3d-printing of microstructures.

%%%%%%%%%%%%%%%%%%%%%%%%%%%%%%%%%%%%%%%%%%%%%%%%%%%%%%%%%%%%%%%%%%%
%
% Bibliography
%
%%%%%%%%%%%%%%%%%%%%%%%%%%%%%%%%%%%%%%%%%%%%%%%%%%%%%%%%%%%%%%%%%%%

%\bibliography{acoustofluidics}


\end{document}